\documentclass[12pt]{article}
\usepackage{amsmath,amssymb,amsthm,amsxtra,overpic,bbm,bm,epsfig,subfigure}
\usepackage{hyperref}
\usepackage{color}
\usepackage{stackengine} 

\begin{document}
\vspace{0.2cm}

\begin{center}
{\Large\bf Slow-roll inflation in $f(R,T)$ gravity with a $RT$ mixing term}\\\vspace{0.2cm}
\end{center}

\vspace{0.2cm}

\begin{center}
{\bf Che-Yu Chen}~$^{a}$~\footnote{Email: b97202056@gmail.com},
\quad
{\bf Yakefu Reyimuaji}~$^{b}$~\footnote{Email: yreyi@hotmail.com},
\quad
{\bf Xinyi Zhang}~$^{c}$~\footnote{Email: zhangxinyi@ihep.ac.cn}
\\
\vspace{0.2cm}
$^a$ {\em \small Institute of Physics, Academia Sinica, Taipei 11529, Taiwan}\\
$^b$ {\em \small School of Physical Science and Technology,
Xinjiang University, Urumqi, Xinjiang 830046, China
}\\
$^c$ {\em \small Institute of High Energy Physics, Chinese Academy of Sciences, Beijing 100049, China}\\
\end{center}

\vspace{0.5cm}

\begin{abstract}
We consider slow-roll inflationary models in a class of modified theories of gravity which contains non-minimal curvature-inflaton couplings, i.e., the $f(R,T)$ gravity, where $R$ is the Ricci scalar and $T$ is the trace of the inflaton energy-momentum tensor. On top of the minimally coupled $T$ that has been widely investigated in the literature, we further include a $RT$ mixing term in the theory. This mixing term introduces non-minimal derivative couplings and plays an important role in inflationary dynamics. Taking chaotic and natural inflation as examples, we find that the predictions for spectral tilt and the tensor-to-scalar ratio are sensitive to the existence of the $RT$ mixing term. In particular, by turning on this mixing term, it is possible to bring chaotic and natural inflation into better agreement with observational data.
\end{abstract}

\vspace{0.5cm}

\section{Introduction}

The standard big bang cosmology in General Relativity (GR), despite its huge success, has some traditional issues, such as the flatness problem and the horizon problem. One promising way for resolving these issues is to introduce an accelerating expansion phase at the early stage of the cosmic evolution, before the onset of the standard big bang cosmology. This early accelerating expansion, known as inflation, also provides a mechanism to seed the current large-scale structure of the universe and the anisotropy that has been accurately measured in the cosmic microwave background (for reviews, see e.g. Refs.~\cite{Lyth:1998xn,Bassett:2005xm,Martin:2013tda}). Among all the inflationary models, the single field slow-roll models are the simplest and are still under active investigation \cite{Chowdhury:2019otk}.

The recently released data from BICEP/Keck \cite{BICEP:2021xfz}, combined with Planck 2018 observations \cite{Planck:2018jri}, put a very strong constraint on one of the inflationary observables, i.e., the tensor-to-scalar ratio $r$. The tightly constrained $r$ already rules out several single field inflationary models, such as the chaotic inflation and the original version of the natural inflation. In fact, to fit into observational constraints, a viable inflationary model has to not only provide enough e-folds, but also predict a small enough tensor-to-scalar ratio. One popular candidate for such viable inflation models is the cosmological $\alpha$-attractor \cite{Kallosh:2013yoa,Kallosh:2022feu}, with the well-known Starobinsky model \cite{Starobinsky:1980te} being its particular case.

In this work, we are going to investigate the possibility of rescuing those observationally disfavored inflationary models by considering gravitational theories beyond GR. In particular, we will consider gravitational theories with non-minimal matter-gravity couplings. In GR, it is assumed that the matter fields are minimally coupled to gravity and the standard conservation equation for the energy-momentum tensor $T_{\mu\nu}$ is satisfied. In the presence of non-minimal matter-gravity couplings, the standard conservation equation may not be valid anymore. The non-conservation of energy-momentum tensor naturally leads to irreversible particle creation processes, or more explicitly, some energy transfer between the gravitational sector and the matte sector. These particle creation processes commonly appear when considering quantum field theory in curved spacetimes. In fact, the presence of non-minimal matter-gravity couplings can be regarded as one potential consequence when considering the semiclassical levels in quantizing gravity \cite{Faraoni:1996rf,Faraoni:2000wk,Dzhunushaliev:2013nea,Dzhunushaliev:2015mva,Lobato:2018vpq,Chen:2021oal}. 

The consideration of non-minimal matter-gravity couplings is also one active research field in the community of modified theories of gravity \cite{Nojiri:2017ncd}. One typical motivation of these types of theories is to explain the late-time accelerating expansion of the universe by resorting to these \textit{dark couplings} between gravity and matter fields, instead of by introducing any exotic dark energy in the matter content. Along this line of thinking, several modified theories of gravity have been proposed and widely studied, such as $f(R,\mathcal{L}_m)$ gravity \cite{Nojiri:2004bi,Allemandi:2005qs,Bertolami:2007gv}, $f(R,T)$ gravity \cite{Harko:2011kv}, and the energy-momentum squared gravity \cite{Roshan:2016mbt}, where $R$, $\mathcal{L}_m$, and $T$ are the Ricci scalar, matter Lagrangian, and the trace of energy-momentum tensor, respectively (for a pedagogical review, see, e.g., Refs.~\cite{Nojiri:2006ri,Harkobook2018}). 

As we have mentioned, non-minimal matter-gravity couplings can be regarded as a potential feature in semiclassical gravity. Therefore, in addition to having an impact on the late-time cosmic expansion, one would expect that these couplings would contribute to the evolution of the early universe, especially to inflationary dynamics. A natural question then arises: Is it possible to reconsider those inflationary models that have been ruled out by observations after assuming that the inflaton fields are non-minimally coupled to gravity? Some indications toward a positive answer have been put forward in the literature \cite{Nozari:2010uu,Boubekeur:2015xza,Pallis:2015mga,Dalianis:2014nwa,Kallosh:2013tua,Reyimuaji:2020goi,Zhang:2021ppy}. 

In this work, we will consider slow-roll inflationary models in the context of $f(R,T)$ gravity. In this case, the trace of the energy-momentum tensor $T$ that appears in the function $f(R,T)$ is assumed to be that of a self-gravitating scalar field \cite{Moraes:2016gpe,Alves:2016iks}. In fact, it has been shown in Ref.~\cite{Gamonal:2020itt} that even if one considers the function $f(R,T)$ a linearly additive function of $R$ and $T$, i.e., a minimally coupled $T$, the chaotic and natural inflationary models are still disfavored by observations (see also Refs.~\cite{Fisher:2019ekh,Bhattacharjee:2020jsf,Deb:2022hna}). In this paper, we will extend the work of Ref.~\cite{Gamonal:2020itt} by introducing two additional terms in the function $f(R,T)$, one is the conformal constant rescaling and the other is a $RT$ mixing term. Unlike the minimally coupled $T$, the $RT$ mixing term directly introduces non-minimal gravity-matter couplings that cannot be recast into a minimal equation by any field redefinitions. This $RT$ mixing term has been shown to provide several interesting consequences both in cosmological \cite{Chakraborty:2012kj,Shabani:2013djy,Sharif:2014ioa,Xu:2016rdf,Sharif:2017lhx,Moraes:2017zgm} and astrophysical scales \cite{Sharif:2012zzd,Bhatti:2017bqa,Zubair:2020poe}. The coupling between matter and geometry generated by this mixing term can play important roles in the early universe. In particular, it may resolve the issue of arrow of time in the context of quantum cosmology \cite{Xu:2016rdf}. In addition, using dynamical analysis \cite{Shabani:2013djy} and model reconstruction approaches \cite{Sharif:2014ioa,Moraes:2017zgm}, respectively, it has been shown that the presence of the mixing term largely enriches the cosmological solutions in the models. Also, the constructed solutions in the presence of the mixing term can be physically justified using Noether symmetry approaches \cite{Sharif:2017lhx}.

In fact, if one assumes that the $T$ in the $RT$ mixing term is contributed by that of the inflaton, the $RT$ mixing term would generate non-minimal derivative coupling terms in the theory. The idea of non-minimally coupling gravity to derivatives of inflaton is actually not new \cite{Amendola:1993uh}. The derivative coupling naturally provides a friction mechanism, helping the satisfaction of slow-roll conditions even if the inflaton potential is steep. Therefore such coupling could significantly enlarge the parameter space in which inflationary attractors exist, validating inflationary models that would otherwise be unsuccessful even if non-minimal coupling terms are introduced \cite{Germani:2010gm}. In addition, it has been shown that derivative couplings could suppress the tensor-to-scalar ratio in various inflationary models and make them compatible with observations \cite{Tsujikawa:2012mk,Nozariand:2015yfi}, as well as modify the reheating scenarios \cite{Sadjadi:2012zp,Ghalee:2013ada,Ema:2015oaa,Dalianis:2016wpu}. In this paper, we will show that inflationary dynamics is indeed very sensitive to the existence of the $RT$ mixing term. We will take chaotic and natural inflationary models as examples and exhibit that the $RT$ mixing term can assist in making these inflationary models compatible with observations.

The paper is organized as follows. In Sec.~\ref{sec:theory}, we introduce the theoretical setup that we are going to consider in this work. In Sec.~\ref{sec:slowrolleq}, we write down the general equations for inflationary models within the slow-roll approximations. Using these equations, in Sec.~\ref{sec:application}, we consider two different inflationary models, i.e., chaotic inflation and natural inflation. Then we calculate the observational predictions for these inflationary models in our theory. We finally conclude in Sec.~\ref{sec:conclude}.

\section{Theoretical setup}\label{sec:theory}
We start with the $f(R,T)$ gravity whose action reads
\begin{equation}
S=\int  d^{4} x\sqrt{-g} \left(\frac{f(R, T)}{2 \kappa}+\mathcal{L}_{m}\right)\,,\label{frt}
\end{equation}
where $\kappa\equiv8\pi G\equiv1/M_\mathrm{pl}^2$ and $g$ is the determinant of the metric $g_{\mu\nu}$. The function $f(R,T)$ is an arbitrary function of the Ricci scalar $R$ and the trace of the energy-momentum tensor $T$. The matter Lagrangian is given by a canonical scalar field $\phi$:
\begin{equation}
\mathcal{L}_{m}=-\frac{1}{2} g^{\mu \nu} \partial_{\mu} \phi \partial_{\nu} \phi-V(\phi)\,,\label{inflatonlagrangian}
\end{equation}
with its potential $V(\phi)$. The trace of the energy-momentum tensor that appears in the function $f(R,T)$ is associated with the scalar field $\phi$, and therefore, for a generic function $f(R,T)$, the theory contains non-minimal curvature-scalar couplings. Note that the reason why we include in addition to the $f(R,T)$ term a standard inflaton Lagrangian \eqref{inflatonlagrangian} in the theory is to ensure that the scenario reduces to a standard minimally coupled inflaton in the GR limit, i.e., when $f(R,T)=R$. Also, in this paper, we will mainly focus on the inflationary regime which is far before reheating. When considering reheating, the standard matter field can be chosen to be either only minimally coupled to gravity, or appear together with the inflaton in the non-minimal $f(R,T)$ term. These two approaches would behave the same during inflation where standard matter field is absent, while they give distinctive results in the reheating phase.

The gravitational equation of the theory can be obtained by varying the action \eqref{frt} with respect to the metric:
\begin{align}
f_RR_{\mu\nu}-\frac{1}{2}fg_{\mu\nu}+\left(g_{\mu\nu}\Box-\nabla_\mu\nabla_\nu\right)f_R=\kappa T_{\mu\nu}-f_T\left(T_{\mu\nu}+\Theta_{\mu\nu}\right)\,,\label{metriceq}
\end{align}
where $f_R\equiv\partial f/\partial R$ and $f_T\equiv\partial f/\partial T$. On the above equation, $\Box\equiv\nabla^\mu\nabla_\mu$ and $\nabla_\mu$ is the covariant derivative. For a canonical scalar field, the energy-momentum tensor is defined by
\begin{equation}
T_{\mu\nu}=\partial_\mu\phi\partial_\nu\phi+g_{\mu\nu}\mathcal{L}_{m}\,.
\end{equation} 
In addition, the tensor $\Theta_{\mu\nu}$ is defined through the variation of the energy-momentum tensor with respect to the metric as follows
\begin{equation}
\Theta_{\mu\nu}\equiv g^{\alpha\beta}\frac{\delta T_{\alpha\beta}}{\delta g^{\mu\nu}}=-\partial_\mu\phi\partial_\nu\phi-T_{\mu\nu}\,.
\end{equation}
After taking the covariant divergence of Eq.~\eqref{metriceq}, one can express the divergence of the energy-momentum tensor as \cite{Alvarenga:2013syu,BarrientosO:2014mys}
\begin{equation}
\left(\kappa-f_T\right)\nabla^\mu T_{\mu\nu}=f_T\left(\nabla^\mu\Theta_{\mu\nu}-\frac{1}{2}\nabla_\nu T\right)+\left(T_{\mu\nu}+\Theta_{\mu\nu}\right)\nabla^\mu f_T\,.\label{conservationofT}
\end{equation}
One sees that in the presence of nonzero $f_T$ on the right-hand side of Eq.~\eqref{conservationofT}, the energy-momentum tensor is in general not conserved.

In the rest of this paper, we are going to consider the following functional form of $f(R,T)$:
\begin{equation}
f(R,T)=R(1+\alpha+\kappa^2\beta T)+\kappa\gamma T\,,\label{ourmodel}
\end{equation}
where $\alpha$, $\beta$, and $\gamma$ are free dimensionless parameters in the theory. This theory can be regarded as a generalization of that in Ref.~\cite{Gamonal:2020itt}, which corresponds to a subset of our model with $\alpha=\beta=0$. By turning on the parameters $\alpha$ and $\beta$, the theory contains a constant conformal rescaling and a direct $RT$ mixing term, respectively. We will show in the next sections that the $RT$ mixing term could have significant impacts on the inflationary observables as compared with those in GR.

\section{Slow-roll inflation}\label{sec:slowrolleq}
In this section, we will write down all the equations that we need for calculating inflationary observables. During inflation, the universe is assumed to be spatially flat, homogeneous, and isotropic. More explicitly, the spacetime can be well-described by the Friedmann-Robertson-Walker (FRW) metric,
\begin{equation}
ds^2=-dt^2+a(t)^2\left(dx^2+dy^2+dz^2\right)
\end{equation}
characterized by a function of cosmic time $t$, namely, the scale factor $a(t)$. The scale factor essentially stands for the size of the universe and its evolution in time can be obtained by solving the field equations of the theory.

Inserting the functional form \eqref{ourmodel} and the FRW metric ansatz into the gravitational equation \eqref{metriceq}, we obtain the following equations of motion:
\begin{align}
H^2(1+\alpha)&=\frac{\kappa}{3}\left[\frac{\dot\phi^2}{2}\left(1+\gamma+18\kappa\beta H^2+12\kappa\beta\dot H\right)+V\left(1+2\gamma+12\kappa\beta H^2\right)\right]\label{firedrt}\nonumber\\&+4\kappa^2\beta H\dot{\phi}V_\phi-2\kappa^2\beta H\dot\phi\ddot{\phi}\,.\\
\frac{\ddot{a}}{a}(1+\alpha)&=-\frac{\kappa}{3}\left[\dot\phi^2\left(1+\gamma+9\kappa\beta H^2+6\kappa\beta\dot{H}-6\kappa\beta V_{\phi\phi}\right)-V\left(1+2\gamma+12\kappa\beta H^2+12\kappa\beta\dot{H}\right)\right]\nonumber\\
&+2\kappa^2\beta H\dot\phi V_\phi-\kappa^2\beta H\dot\phi\ddot\phi+2\kappa^2\beta\ddot\phi V_\phi-\kappa^2\beta\ddot\phi^2-\kappa^2\beta\dot\phi\dddot\phi\,,\label{raychoumodi}
\end{align}
where $V_\phi\equiv dV/d\phi$, $V_{\phi\phi}\equiv d^2V/d\phi^2$, and $H\equiv\dot{a}/a$ is the Hubble function. The dot denotes the derivatives with respect to $t$. Furthermore, the scalar field equation, which can be obtained by varying the action \eqref{frt} with respect to the scalar field $\phi$, can be written as
\begin{align}
\ddot{\phi}\left(1+\gamma+12\kappa\beta H^2+6\kappa\beta\dot H\right)&+3H\dot\phi\left(1+\gamma+10\kappa\beta H^2+8\kappa\beta\dot H\right)\nonumber\\&+V_\phi\left(1+2\gamma+24\kappa\beta H^2+12\kappa\beta\dot{H}\right)+6\kappa\beta\dot\phi\frac{\dddot{a}}{a}=0\,.\label{kgeqet}
\end{align}
The scalar field equation \eqref{kgeqet} can also be obtained using Eq.~\eqref{conservationofT}. It can be seen that the above equations \eqref{firedrt}, \eqref{raychoumodi}, and \eqref{kgeqet} reduce to those of GR minimally coupled to a scalar field when $\alpha=\beta=\gamma=0$. We would like to emphasize that a non-zero $\beta$ introduces terms with derivatives higher than two in the equations, as one can see from the presence of the terms containing $\dddot{\phi}$, $\dddot{a}$, $\ddot{\phi}^2$, and so forth. This is expected because the $RT$ mixing term in the action naturally introduces derivative couplings in the theory.

One common step, which can be regarded as a trick, when considering inflationary models in modified theories of gravity is to recast the theory into its Einstein frame by adopting conformal transformations and field redefinitions. One typical example is the Starobinsky inflationary model, which is originated from an action with $R^2$ term in the Jordan frame. The presence of the $R^2$ term is associated with a dynamical scalar degree of freedom, which can be expressed as a scalar field minimally coupled to the Einstein-Hilbert action after recasting the theory into the Einstein frame. In the Einstein frame, the calculations of the inflationary observables are straightforward, and this method can be applied even when the scalar field is non-minimally coupled to gravity. However, it has been shown in Ref.~\cite{Amendola:1993uh} that, in general, the Einstein frame representation does not exist when non-minimally derivative couplings are included in the theory. Therefore, the calculations of inflationary observables based on Einstein frame representation do not work for our theory in the presence of a non-zero $\beta$.

However, if one considers strictly the slow-roll approximations and neglects higher-order derivative terms in the equations before going to the Einstein frame, it is still possible to calculate the inflationary observables by transforming the system to the Einstein frame. The typical assumptions of slow-roll inflationary models are first that the scalar field is slowly rolling on its potential, and each quantity, such as $H$ and $\phi$, is changing quasi-statically. Therefore, in the slow-roll approximation, we assume $\dot\phi^2\ll V(\phi)$, $|\dddot{\phi}|\ll |H\ddot{\phi}|\ll|H^2\dot{\phi}|$, and $|\ddot{H}|\ll|H\dot{H}|\ll|H^3|$. The modified Klein-Gordon equation \eqref{kgeqet} and the modified Friedmann equation \eqref{firedrt} can be approximated as
\begin{equation}
3H\dot\phi\left(1+\gamma+12\kappa\beta H^2\right)+V_\phi\left(1+2\gamma+24\kappa\beta H^2\right)\approx0\,,\label{slkg}
\end{equation}
and
\begin{equation}
H^2\left(1+\alpha\right)\approx\frac{\kappa}{3}V\left(1+2\gamma+12\kappa\beta H^2\right)\,,\label{slowfriedrt}
\end{equation}
respectively. Furthermore, combining Eq.~\eqref{raychoumodi} and the modified Friedmann equation, one obtains
\begin{equation}
\dot{H}\left(1+\alpha-4\kappa^2\beta V\right)\approx-\kappa\frac{\dot{\phi}^2}{2}\left(1+\gamma+12\kappa\beta H^2\right)-2\kappa^2\beta H\dot\phi V_\phi\,.\label{sleq3}
\end{equation}

Our next step is to rewrite the above equations \eqref{slkg}, \eqref{slowfriedrt}, and \eqref{sleq3} into their Einstein frame representations. To achieve this, we define three functions $\Omega_1(\phi)$, $\Omega_2(\phi)$, and $\Omega_3(\phi)$. The first function $\Omega_1(\phi)$ stands for a conformal factor that transforms the metric $g_{\mu\nu}$ to an auxiliary metric $\tilde{g}_{\mu\nu}$:
\begin{equation}
\tilde{g}_{\mu\nu}=\Omega_1(\phi) g_{\mu\nu}\,.
\end{equation}
More explicitly, in the FRW ansatz, the two line elements are related as follows 
\begin{align}
\tilde{g}_{\mu\nu}dx^\mu dx^\nu&=-d{\tilde{t}}^2+\tilde{a}(\tilde{t})^2dx_idx^i\nonumber\\&\equiv\Omega_1\left(-dt^2+a(t)^2dx_idx^i\right)\,,
\end{align}
where $i$ runs from $i=1,2,3$. The second function $\Omega_2(\phi)$ is used to define an auxiliary scalar field $\chi$, such that
\begin{equation}
\left(\frac{d\chi}{d\phi}\right)^2=\Omega_2(\phi)\,.
\end{equation}
The third function $\Omega_3(\phi)$ is used to rescale the potential:
\begin{equation}
\tilde{V}(\chi(\phi))=\Omega_3(\phi) V(\phi)\,.
\end{equation}
In order to recast the slow-roll equations into their Einstein frame representations, we shall choose the three functions properly such that Eqs.~\eqref{slkg}, \eqref{slowfriedrt}, and \eqref{sleq3} can be written as
\begin{align}
&3\tilde{H}\frac{d\chi}{d\tilde{t}}+\tilde{V}_\chi\approx0\,,\label{slkgre}\\
&\tilde{H}^2\approx\frac{\kappa}{3}\tilde{V}\,,\label{conffiredsl}\\
&\frac{d\tilde{H}}{d\tilde{t}}\approx-\frac{\kappa}{2}\left(\frac{d\chi}{d\tilde{t}}\right)^2\,,\label{htsl}
\end{align}
where $\tilde{V}_\chi\equiv d\tilde{V}/d\chi$ and $\tilde{H}=(d\tilde{a}/d\tilde{t})/\tilde{a}$ is the Hubble function defined by the auxiliary metric. This can be done by choosing 
\begin{align}
\Omega_1&=1+\alpha-4\kappa^2\beta V\,,\\
\Omega_2&=\frac{\left(1+\alpha\right)\left(1+\gamma\right)+4\beta\gamma \kappa^2 V}{\left(1+\alpha-4\beta\kappa^2 V\right)^2}\,,\\
\Omega_3&=\frac{1+2\gamma}{\left(1+\alpha-4\beta\kappa^2 V\right)^2}\,.
\end{align}

In the Einstein frame in which the dynamical equations are given by Eqs.~\eqref{slkgre}, \eqref{conffiredsl}, and \eqref{htsl}, the inflationary observables can be calculated by first defining the slow-roll parameters 
\begin{align}
&\epsilon_{\tilde{V}} = \displaystyle \frac{1}{2\kappa} \left(\frac{\tilde{V}_{\chi}}{\tilde{V}} \right)^2\,, \label{eq:epsilonGR}\\
&\eta_{\tilde{V}} = \displaystyle \frac{1}{\kappa}\frac{\tilde{V}_{\chi\chi}}{\tilde{V}}\,,\label{eq:etaGR}
\end{align}
where $\tilde{V}_{\chi\chi}\equiv d^2\tilde{V}/d\chi^2$. The end of inflation is practically determined by the time when either $\epsilon_{\tilde{V}}$ or $|\eta_{\tilde{V}}|$ approaches unity. A successful inflationary model that is consistent with observations has to last for long enough e-folds. The e-folding number $N$ is defined by
\begin{equation}
N\equiv\int_{\tilde{a}_i}^{\tilde{a}_f}\frac{d\tilde{a}}{\tilde{a}}=\int_{\chi_i}^{\chi_f}\frac{\tilde{H}}{d\chi/d\tilde{t}}d\chi\,,
\end{equation}
with the subscripts $i$ and $f$ denoting the beginning and the end of the inflation, respectively. Once the time of the end of inflation is determined, one can calculate the inflaton and the field values at the horizon crossing by counting backward the e-folding number. Finally, the inflationary observables, namely, the spectral tilt and the tensor-to-scalar ratio, which are calculated at the horizon crossing, can be obtained via
\begin{align}
n_s =1-6\epsilon_{\tilde{V}}+2\eta_{\tilde{V}}\;,\quad
r =16\epsilon_{\tilde{V}}\,.\label{eq:nsrGR}
\end{align}

In the next section, we will consider two specific inflationary potentials, i.e., the chaotic inflation and the natural inflation, and calculate their inflationary observables using Eq.~\eqref{eq:nsrGR}, which are tightly based on the mapping from the Jordan frame (Eqs.~\eqref{slkg}, \eqref{slowfriedrt}, and \eqref{sleq3}) to the Einstein frame (Eqs.~\eqref{slkgre}, \eqref{conffiredsl}, and \eqref{htsl}). However, we should emphasize that this mapping is in general not valid for theories with non-minimal derivative couplings, in particular in the presence of a non-zero $\beta$. Our results here strongly rely on the validity of the slow-roll approximations and on the requirement that the slow-roll approximation terminates abruptly enough near the end of the inflation. Although the two requirements can be confirmed according to our numerical calculations, the inflationary observables that are obtained using this mapping should not be treated as accurate predictions of the model. However, we expect that the results should still be able to capture the qualitative features of the model that are generated by the $RT$ mixing term.

\section{Application to inflationary models}
\label{sec:application}
\subsection{Chaotic inflation}

\begin{figure}[t!]
\centering
\includegraphics[width=0.96\textwidth]{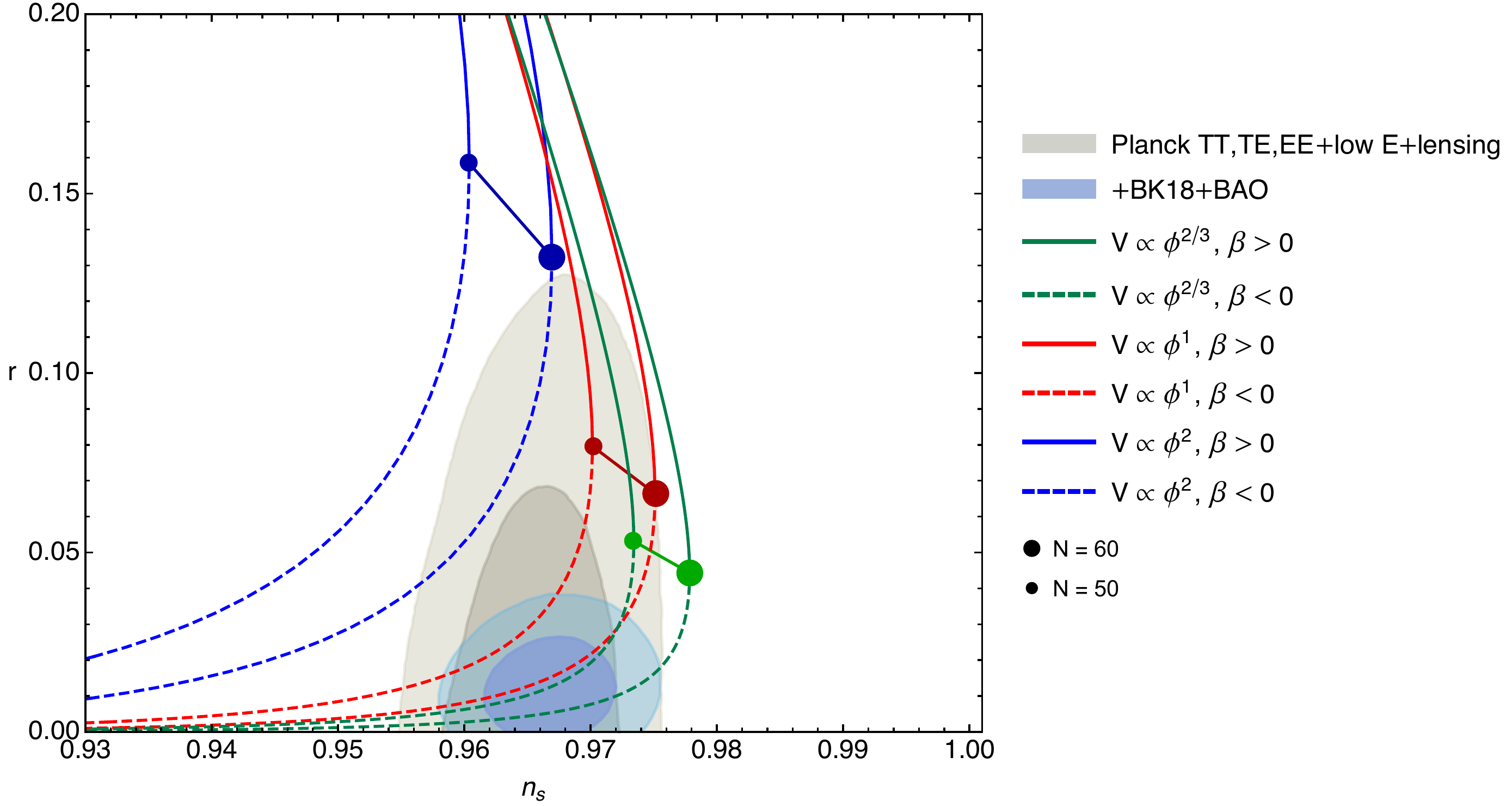}
\caption{The spectral tilt $n_s$ and tensor-to-scalar ratio $r$ for chaotic inflationary model in the $f(R,T)$ theory \eqref{ourmodel}. The shaded regions show the contours of the latest Planck 2018 \cite{Planck:2018jri} with (blue) and without (gray) the BICEP/Keck data \cite{BICEP:2021xfz}. In this figure, we choose $\alpha=\gamma=0$ and focus only on the effects of changing $\beta$, namely, the $RT$ coupling coefficient. The blue, red, and green curves represent $n=2$, $1$, and $2/3$, respectively. The results in GR are shown by the lines connecting the two circles, ranging from $N=50$ to $N=60$. The ranges of $\beta$ are $[-0.0011,0.0002]$, $[-0.031,0.018]$, and $[-0.082,0.0571]$, respectively, for $n=2$, $1$, and $2/3$.}
\label{fig:model_chaotic}
\end{figure}
We first consider the chaotic inflation model \cite{Linde:1983rmu}, where the potential takes the power-law form
\begin{align}
V(\phi)=\lambda M_\mathrm{pl}^4 \left( \displaystyle \frac{\phi}{M_\mathrm{pl}} \right)^n \,,\label{chaoticpo}
\end{align}
where $n$ is the power index and $\lambda$ is a dimensionless coupling constant.  Chaotic inflation is a typical large-field inflation model where the inflaton traverses super-Planckian distances in field space. The word ``chaotic” refers to the initial state of the universe. It possesses one of the simplest forms of single-field potential thus is considered as a prototype model and received a lot of studies. In GR, the chaotic inflationary model generically predicts a relatively large tensor-to-scalar ratio. In particular, for $n\ge2$, the chaotic inflationary model is strongly disfavored by Planck 2018~\cite{Planck:2018jri}. In this subsection, we will consider the potential \eqref{chaoticpo} in our model \eqref{ourmodel} and see how the three parameters $\alpha$, $\beta$, and $\gamma$ alter the inflationary observables.

First, we observe that the parameter $\beta$, namely, the $RT$ coupling term, is necessary for chaotic inflation in order to get non-trivial results. This can be exhibited by expanding the field values and the slow-roll parameters in terms of $\beta$. We find that up to the zeroth-order of $\beta$, the auxiliary scalar field $\chi$, the slow-roll parameters, and the derivative of the e-folding number with respect to $\chi$, can be expressed as
\begin{align}
\chi(\phi)&\approx\sqrt{\frac{1+\gamma}{1+\alpha}}\phi+\mathcal{O(\beta)}\,,\\
\epsilon_{\tilde{V}}\approx\frac{n^2}{2\kappa\chi^2}+\mathcal{O(\beta)}\,,\quad
\eta_{\tilde{V}}&\approx\frac{n\left(n-1\right)}{\kappa\chi^2}+\mathcal{O(\beta)}\,,\quad
\frac{dN}{d\chi}\approx\frac{\kappa\chi}{n}+\mathcal{O(\beta)}\,.
\end{align}  
One can clearly see that if $\beta$ is turned off, the expressions of the slow-roll parameters and the e-folding number in terms of $\chi$ do not depend on $\{\alpha,\gamma\}$. Once the e-folding number is fixed by observations, the slow-roll parameters at the horizon crossing are determined, so are the inflationary observables $n_s$ and $r$. Therefore, if $\beta=0$, the predicted curves on the $n_s$-$r$ plane for the chaotic inflationary model are the same as those in GR. This is consistent with the results found in Ref.~\cite{Gamonal:2020itt}, in which the author found that the $n_s$-$r$ curves for a subclass of our model with $\alpha=\beta=0$ are the same as those in GR. Therefore, the attempt to alter the $n_s$-$r$ curves and make them consistent with Planck results requires a non-zero $\beta$.

In Figure~\ref{fig:model_chaotic}, we fix $\alpha=\gamma=0$, and show the $n_s$-$r$ predictions of the model \eqref{ourmodel} when varying $\beta$. The shaded regions show the contours of the latest Planck 2018 data \cite{Planck:2018jri} with (blue) and without (gray) the BICEP/Keck results \cite{BICEP:2021xfz}. The blue, red, and green curves represent $n=2$, $n=1$, and $n=2/3$, respectively. The solid and dashed curves correspond to a positive and negative $\beta$, respectively. The solid and dashed curves are joined by the colored circles ($\beta=0$), which represent the GR results with $N=50$ and $60$. The ranges of $\beta$ for each colored curve are detailed in the caption. One clearly sees that the $n_s$-$r$ predictions are very sensitive to $\beta$. In particular, a negative $\beta$ may be able to give $n_s$-$r$ predictions consistent with the Planck contours.

\begin{figure}[t!]
\centering
\includegraphics[width=0.45\textwidth]{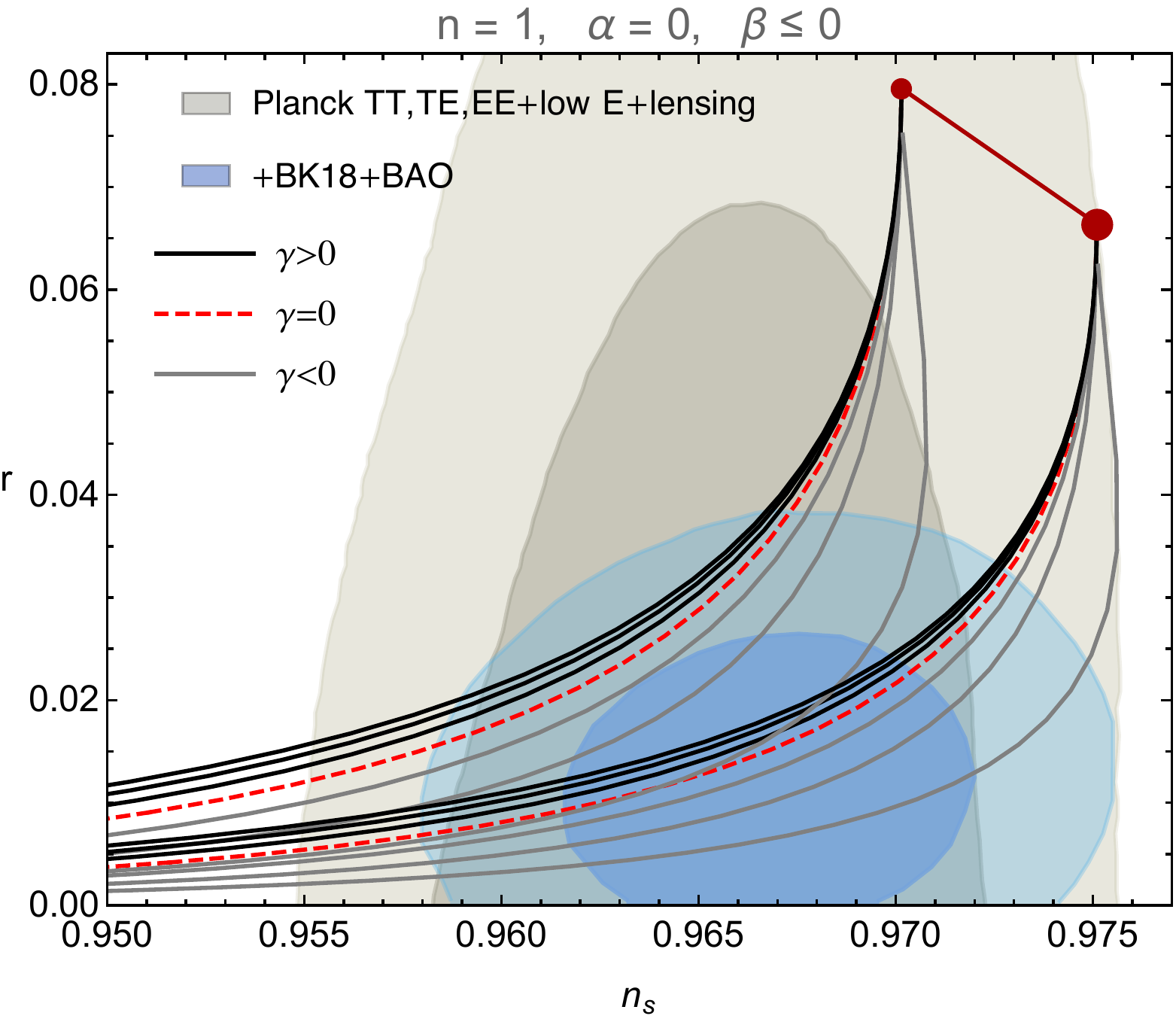}
\includegraphics[width=0.45\textwidth]{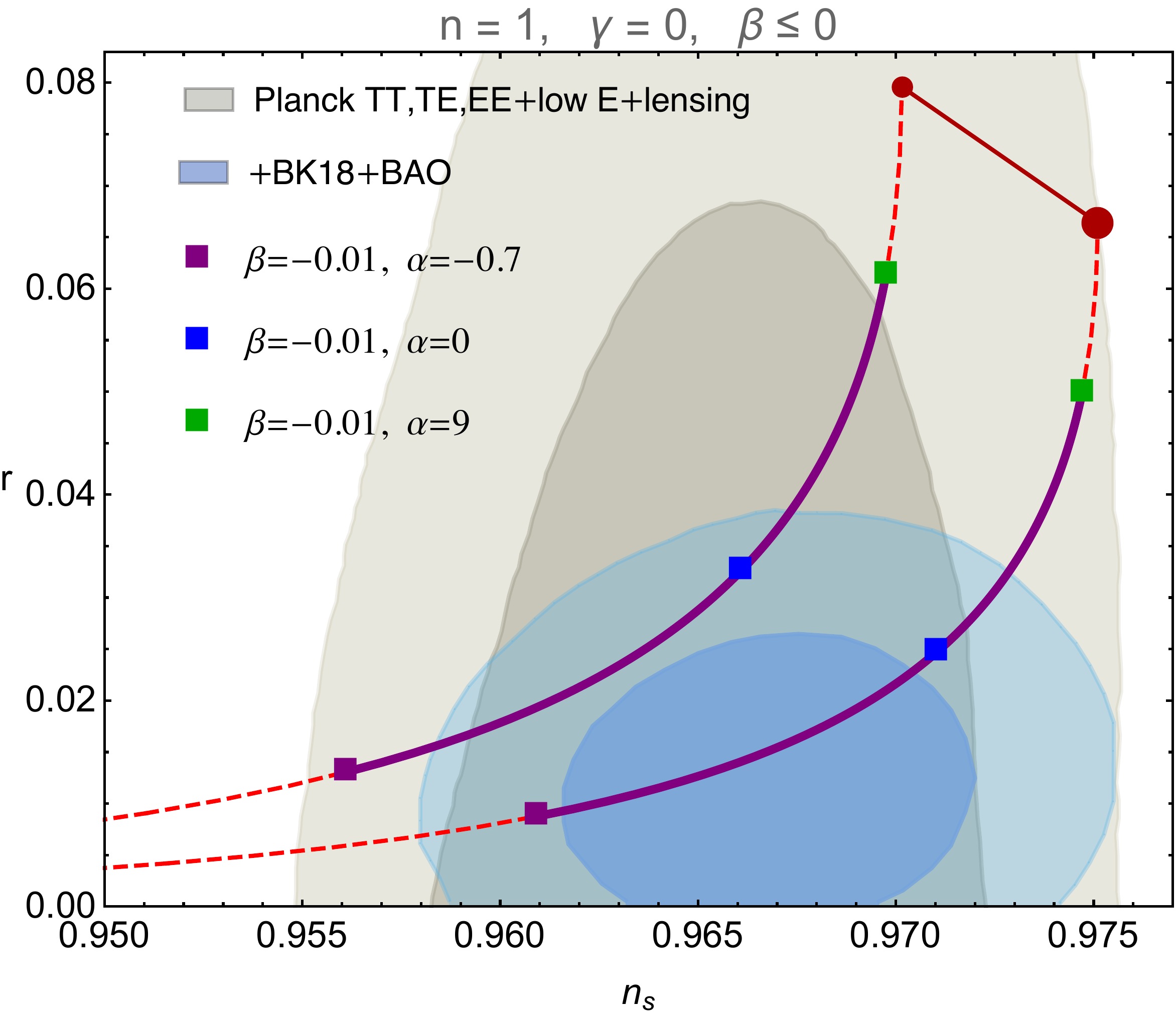}
\caption{The $n_s$-$r$ curves for chaotic inflation with $n=1$ in the $f(R,T)$ theory \eqref{ourmodel}. Left: We assume $\alpha=0$ and focus on $\beta\le0$. For a fixed e-folding number $N$, the black curves (from top to bottom), the red dashed curve, and the gray curves (from top to bottom) correspond to $\gamma=0.9$, $0.6$, $0.3$, $0$, $-0.3$, $-0.6$, and $-0.9$, respectively. Right: We fix $\gamma=0$ and focus on $\beta\le0$. The red dashed curves show the results for $\alpha=0$ and $\beta\le0$. On the purple curves, we fix $\beta=-0.01$ and vary $\alpha$ in the range $-0.7\le\alpha\le9$. Strong degeneracy between the parameters $\alpha$ and $\beta$ can be clearly seen.}
\label{fig:model_chaotic2}
\end{figure}

In Figure~\ref{fig:model_chaotic2}, we exhibit the $n_s$-$r$ results in the parameter space $\alpha$, $\beta$, and $\gamma$. We take the chaotic inflationary model with $n=1$ for demonstration. On  the left panel, we fix $\alpha=0$, and focus on $\beta\le0$. The black and gray curves represent a positive and a negative $\gamma$, respectively (the assigned values of $\gamma$ are detailed in the caption). It can be seen that, as compared with the red dashed curves, a negative $\gamma$ can further bring the $n_s$-$r$ results into better agreement with observations.

On the right panel of Figure~\ref{fig:model_chaotic2}, we fix $\gamma=0$ and again, focus on $\beta\le0$. The purple curves represent the results with $\beta=-0.01$ and a varying $\alpha$ in the range $-0.7\le\alpha\le9$. It can be seen that when $\gamma=0$, there is a strong degeneracy between the parameters $\alpha$ and $\beta$ on the $n_s$-$r$ plane. Namely, one can obtain the predictions of $n_s$-$r$ pair on the red dashed curves with a different value of $\beta$ by just changing $\alpha$. This degeneracy cannot be straightforwardly shown analytically. Therefore, we exhibit this degeneracy numerically in this figure.

Before closing this subsection, we would like to also mention that when $\beta=0$, the $n_s$ and $r$ predictions in Figure~\ref{fig:model_chaotic2} do not depend on $\alpha$ nor $\gamma$. This is consistent with our discussion before that to alter the $n_s$-$r$ curves for chaotic inflation, a non-zero $\beta$ is necessary. In fact, the $n_s$-$r$ curves for chaotic inflation are quite sensitive to $\beta$, but not very sensitive to the other two parameters $\{\alpha,\gamma\}$.

\subsection{Natural inflation}

\begin{figure}[t!]
\centering
\includegraphics[width=0.96\textwidth]{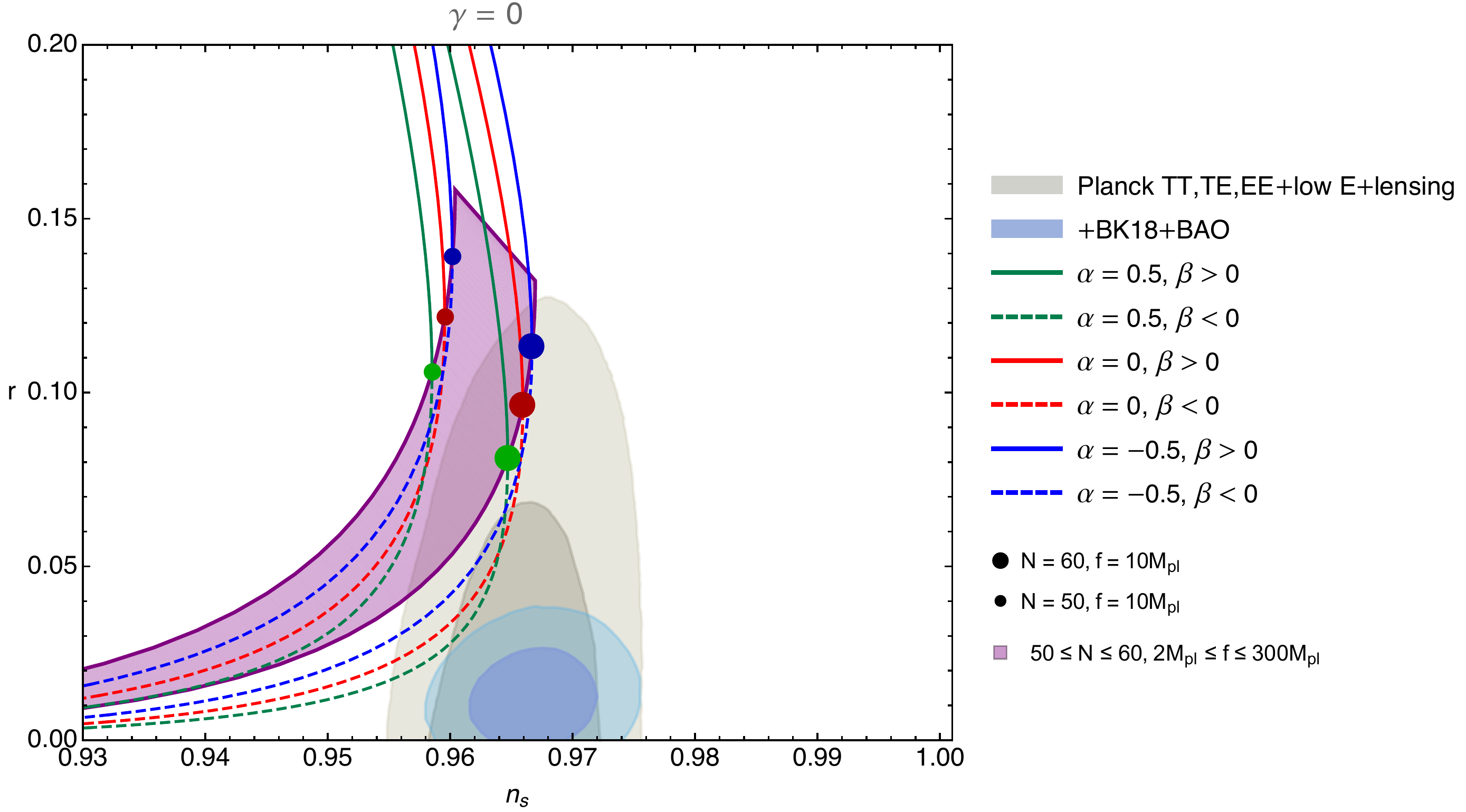}
\caption{The $n_s$-$r$ curves for the natural inflationary model in the $f(R,T)$ theory \eqref{ourmodel} with $\gamma=0$. The blue and gray shaded regions show the contours of the latest Planck 2018 \cite{Planck:2018jri} with and without the BICEP/Keck data \cite{BICEP:2021xfz}, respectively. The shaded purple region shows the $n_s$-$r$ predictions for natural inflation in the parameter space $50\le N\le60$ and $2M_\mathrm{pl}\le f\le300M_\mathrm{pl}$ in GR. Fixing $f=10M_\mathrm{pl}$ and assuming $N=50$ or $60$, we vary $\alpha$ and $\beta$. The solid (dashed) curves correspond to a positive (negative) $\beta$. The green, red, and blue curves represent $\alpha=0.5$, $0$, and $-0.5$, respectively. Note that when $\alpha\ne0$, the $n_s$-$r$ curves still cross the boundary of the shaded purple region at $\beta=0$ (see the blue and green circles).}
\label{fig:model_natural}
\end{figure}



In this subsection, we consider Natural inflation \cite{Freese:1990rb,Adams:1992bn}, in which the inflaton can be interpreted as an axion-like particle moving on a potential of the form
\begin{align}
V=\Lambda^4 \left(1+\cos\phi/f \right)\,,
\end{align}
where $\Lambda$ is the inflationary energy scale, and $f$ is the decay constant{\footnote{Not to confuse it with the functional $f(R,T)$ in the theory. At this point, we have assumed a specific function for $f(R,T)$ as in Eq.~\eqref{ourmodel}}}. The axion-like particle arises naturally whenever there is a global symmetry spontaneously breaking and the flatness of the potential gets naturally protected by the shift symmetry. In GR, the natural inflationary model is disfavored by the Planck observation because it predicts a relatively large $r$ when $f\gtrsim 10M_\mathrm{pl}$, and a relatively small $n_s$ when $f\sim M_\mathrm{pl}$. See the shaded purple region in Figure~\ref{fig:model_natural}.

As what we have done for chaotic inflation models, we will first show that the parameter $\beta$, which represents the direct $RT$ mixing term in the theory, should be present in order to obtain non-trivial results for the $n_s$-$r$ predictions. This can be exhibited by assuming $\beta=0$. In this case, the auxiliary scalar field $\chi$ and the slow-roll parameters for natural inflation in terms of $\phi$ can be expressed as
\begin{align}
\chi(\phi)&=\sqrt{\frac{1+\gamma}{1+\alpha}}\phi\,,\\
\epsilon_{\tilde{V}}=\frac{(1+\alpha)\sin^2{(\phi/f)}}{2\kappa f^2(1+\gamma)\left(1+\cos(\phi/f)\right)^2}\,,&\qquad \eta_{\tilde{V}}=-\frac{(1+\alpha)\cos(\phi/f)}{\kappa f^2(1+\gamma)\left(1+\cos(\phi/f)\right)}\,.
\end{align}
If one rescales the decay constant as
\begin{equation}
\tilde{f}\equiv \sqrt{\frac{1+\gamma}{1+\alpha}}f\,,
\end{equation} 
such that $\phi/f=\chi/\tilde{f}$, the slow-roll parameters can be rewritten as
\begin{equation}
\epsilon_{\tilde{V}}=\frac{1}{2\kappa\tilde{f}^2}\left[\frac{\sin(\chi/\tilde{f})}{1+\cos(\chi/\tilde{f})}\right]^2\,,\qquad\eta_{\tilde{V}}=-\frac{1}{\kappa\tilde{f}^2}\frac{\cos(\chi/\tilde{f})}{1+\cos(\chi/\tilde{f})}\,.\label{slowrollrescalef}
\end{equation}
Therefore, the slow-roll parameters reduce to those in GR upon a rescaling of the decay constant $f$. Because the spectral tilt $n_s$ and the tensor-to-scalar ratio $r$ are solely defined by the slow-roll parameters as given in Eq.~\eqref{eq:nsrGR}, changing the parameters $\alpha$ and $\gamma$ only gives rise to the $n_s$-$r$ predictions that can completely be obtained in GR by tuning the decay constant $f$. Because the whole $n_s$-$r$ band of natural inflation in GR is completely disfavored by the Planck observation, assuming $\beta=0$ and only changing $\alpha$ and $\gamma$ do not help to rescue the natural inflationary model. In the rest of this section, we will then keep turning on the value of $\beta$, and see how the $n_s$-$r$ predictions vary in different regions of the parameter space.

First, in Figure~\ref{fig:model_natural}, we fix $\gamma=0$ and see how the $n_s$-$r$ predictions of natural inflation are altered by the parameters $\alpha$ and $\beta$. Fixing the values of $f=10M_\mathrm{pl}$ and $N=50$ or $60$, we vary the values of $\alpha$ and $\beta$. The solid curves represent a positive $\beta$, while the dashed curves represent a negative $\beta$. On the other hand, the green, red, and blue curves correspond to $\alpha=0.5$, $0$, and $-0.5$, respectively. One can notice that when $\alpha\ne0$, the $n_s$-$r$ curves still cross the boundary of the purple shaded region at $\beta=0$. This is consistent with what we have shown in Eq.~\eqref{slowrollrescalef} that when $\beta=0$, there is a degeneracy between $\alpha$, $\gamma$, and the decay constant $f$.

The most important observation from Figure~\ref{fig:model_natural} is that for the chosen value of $f$ here, the models with a negative $\beta$ and a positive $\alpha$ can bring the $n_s$-$r$ curves slightly downward. However, the predictions are still disfavored by the Planck 2018 data once the latest BICEP/Keck \cite{BICEP:2021xfz} results are taken into account. Therefore, later we shall see whether the curves can be brought further downward when turning on the parameter $\gamma$.

\begin{figure}[t!]
\centering
\includegraphics[width=0.96\textwidth]{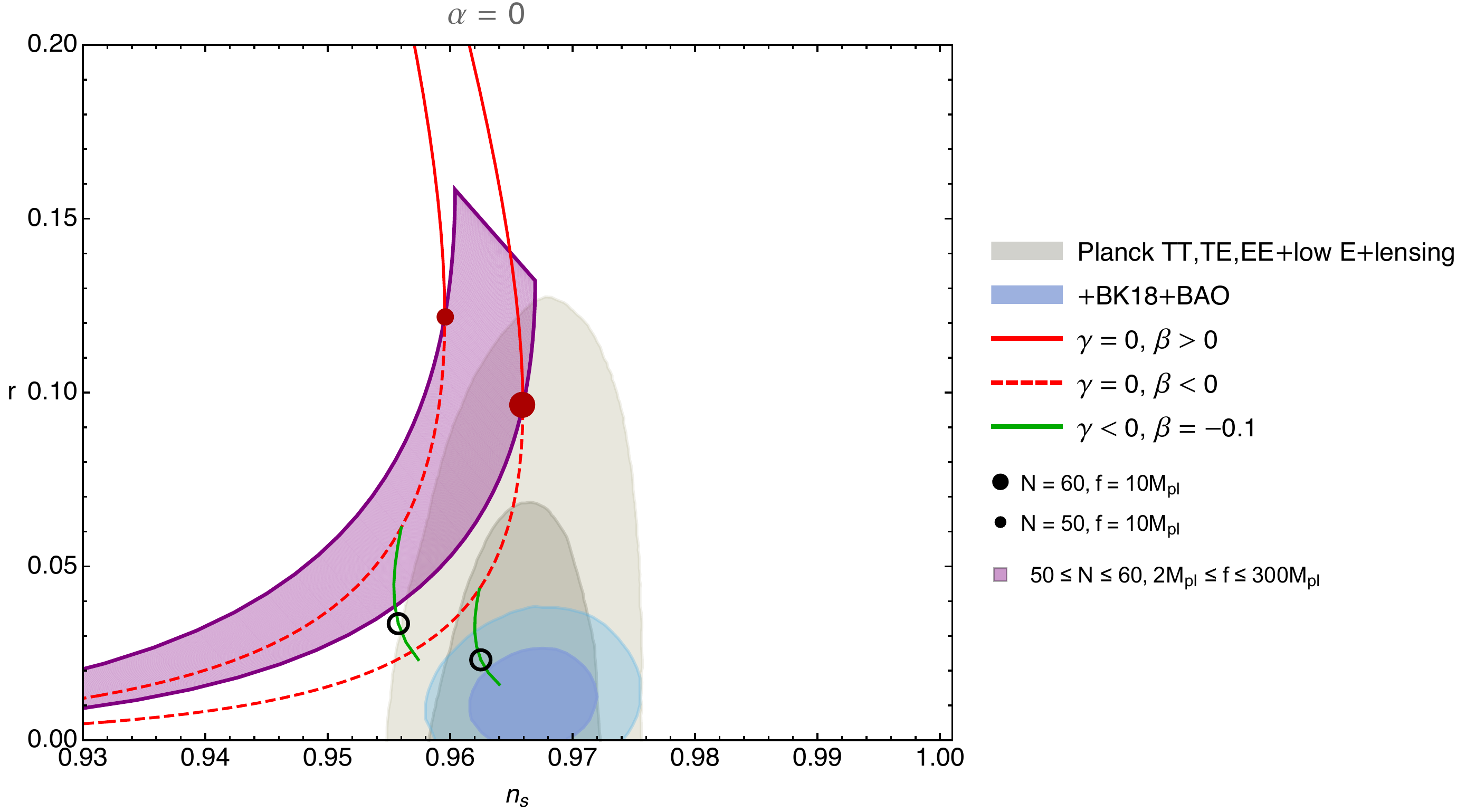}
\caption{The $n_s$-$r$ curves for the natural inflationary model in the $f(R,T)$ theory \eqref{ourmodel} with $\alpha=0$. The shaded purple region represents the parameter space $50\le N\le60$ and $2M_\mathrm{pl}\le f\le300M_\mathrm{pl}$ in GR. Fixing $f=10M_\mathrm{pl}$ and $\gamma=0$, the solid red (dashed red) curves represent $\beta>0$ ($\beta<0$). Starting with $\beta=-0.1$, the green curves show the results when varying negative values of $\gamma$. The open circles denote $\gamma=-0.5$ on which the potential $\tilde{V}$ identically vanishes.}
\label{fig:model_natural_gamma}
\end{figure}

Now, we fix $\alpha=0$ and vary the values of $\beta$ and $\gamma$, to see how the predictions of natural inflationary models are altered. The results are shown in Figure~\ref{fig:model_natural_gamma}. In this figure, we first fix $\alpha=\gamma=0$ and $f=10M_\mathrm{pl}$, then vary the value of $\beta$. The solid red and dashed red curves correspond to a positive and a negative $\beta$, respectively. Then, at $\beta=-0.1$, we start to vary the value of $\gamma$ to its negative direction, as shown by the green curves. One can see that a negative $\gamma$ can bring the $n_s$-$r$ curves further into the Planck 2018 contours. We would like to also emphasize that when $\gamma=-0.5$, the function $\Omega_3$ identically vanishes. Therefore, the potential $\tilde V$ also vanishes. Although the $n_s$ and $r$ can still be defined because the constant factor $1+2\gamma$ can be canceled in the mathematical expressions of the two slow-roll parameters, the predictions of $n_s$ and $r$ are not reliable in this specific case. We highlight this specific case using the open circles in Figure~\ref{fig:model_natural_gamma}.

Finally, we would like to discuss the possibility of having an observationally consistent natural inflationary model with $f\sim M_\mathrm{pl}$ in the theory \eqref{ourmodel}. In table \ref{summaryf1}, we consider $f=M_\mathrm{pl}$ and $N=60$, then summarize the $n_s$-$r$ predictions in the model \eqref{ourmodel} within some regions of the parameter space. According to this table, it can be seen that by properly choosing the parameters $\alpha$, $\beta$, and $\gamma$, the natural inflationary model with $f\sim M_\mathrm{pl}$ could give an extremely small $r$, and a value of $n_s$ very close to the edge of the Planck $1\sigma$ region. One can also notice that all the parameters $\alpha$, $\beta$, and $\gamma$ shown in table \ref{summaryf1} have negative values. We are not excluding the possibilities of having observationally consistent $n_s$-$r$ predictions for natural inflationary models with $f\sim M_\mathrm{pl}$ in the parameter space where at least one of the three parameters $\alpha$, $\beta$, and $\gamma$ is positive. However, it seems that having all these parameters negative is much easier to obtain consistent predictions of $n_s$-$r$ pair.

\begin{table*}
 \begin{center}
  \begin{tabular}{||c|c|c|c|c||} 
 \cline{1-4}
   \hline\hline
   $\alpha$ & $\beta$ & $\gamma$ & $n_s$ & $r$   \\ \hline\hline $-0.19$ & $-0.1$ & $-0.001$ & $0.954206$ & $1.1081\times 10^{-5}$\\ \hline
   $-0.19$ & $-0.1$ & $-0.00186209$ & $0.954208$ & $1.10791\times 10^{-5}$\\ \hline
   $-0.19$ & $-0.1$ & $-0.00346737$ & $0.954212$ & $1.10755\times 10^{-5}$\\ \hline
   $-0.19$ & $-0.1$ & $-0.00645654$ & $0.95422$ & $1.10689\times 10^{-5}$\\ \hline
   $-0.19$ & $-0.1$ & $-0.0120226$ & $0.954234$ & $1.10567\times 10^{-5}$\\ \hline
   $-0.19$ & $-0.1$ & $-0.00223872$ & $0.95426$ & $1.10343\times 10^{-5}$\\ \hline
       \end{tabular}
  \caption{This table summarizes the $n_s$-$r$ predictions of the natural inflation with $f=M_\mathrm{pl}$ in the modified gravity \eqref{ourmodel}. We fix the e-folding number to be $N=60$. In some regions of the parameter space of mostly negative $\{\alpha,\beta,\gamma\}$, an extremely small $r\approx 10^{-5}$ is attainable.}
    \label{summaryf1}
 \end{center}
\end{table*}

There are several other attempts to bring natural inflation into agreement with Planck results, such as with a warm dissipative effect~\cite{Reyimuaji:2020bkm}, or other forms of modified gravity, e.g., in a UV complete quadratic gravity~\cite{Salvio:2019wcp,Salvio:2021lka,Salvio:2022mld}, a non-minimal coupling to Ricci scalar~\cite{Reyimuaji:2020goi}, $f(R,T)$ gravity~\cite{Gamonal:2020itt}. It is worth noticing that the simple form of $f(R,T)$ considered in Ref.~\cite{Gamonal:2020itt} cannot modify natural inflation $n_s$-$r$ prediction, as has also been confirmed in Eq.~\eqref{slowrollrescalef} above.

\section{Conclusions}\label{sec:conclude}

In this paper, we investigate slow-roll inflationary models in a class of non-minimally coupled theories of gravity, namely, the $f(R,T)$ gravity. We assume that the trace of the energy-momentum tensor $T$ appearing in the gravitational action stands for that of the inflaton, giving rise to non-minimal curvature-inflaton couplings in the theory. In particular, in addition to a constant conformal rescaling (parameterized by $\alpha$) and a minimally coupling term (parameterized by $\gamma$), which have been widely considered in the literature, in this work we include a $RT$ mixing term (parameterized by $\beta$) and focus on its effects on the inflationary dynamics. 

The $RT$ mixing term directly introduces non-minimal derivative couplings, and higher-order derivative terms naturally appear in the equations of motion. By assuming strictly that the inflaton slowly rolls on its potential, we neglect the contributions of those higher-order derivative terms and manage to transform the equations of motion into their Einstein frame representations, from which the spectral tilt $n_s$ and the tensor-to-scalar ratio $r$ can be straightforwardly calculated.

We find that for chaotic and natural inflationary models, the $RT$ mixing term plays a significant role in determining the $n_s$-$r$ pair. We show that if the $RT$ mixing term is turned off, namely, $\beta=0$, the other parameters $\alpha$ and $\gamma$ do not alter the $n_s$-$r$ curves. This agrees with what has been shown in the literature. In fact, in order to bring the $n_s$-$r$ curves into better agreement with observational data, a negative $\beta$ seems to be required. In addition, although a negative $\beta$ alone already works quite well in rescuing the chaotic inflationary models (Figure~\ref{fig:model_chaotic}), including also a negative $\gamma$ can further suppress the tensor-to-scalar ratio and make the predictions more consistent with observations (Figure~\ref{fig:model_chaotic2}). The parameter $\alpha$, on the other hand, is not efficient enough to suppress $r$ .

As for the natural inflation, we find that including jointly $\alpha$ and $\beta$ is not able to suppress $r$ by an enough amount to fit in with observations (Figure~\ref{fig:model_natural}). In order to rescue natural inflation in this theory, a negative $\gamma$ is necessary, as one can see from the green curve in Figure~\ref{fig:model_natural_gamma}. Furthermore, we show that by choosing properly the regions of the parameter space, it is possible to rescue the natural inflation with a decay constant $f\sim M_\mathrm{pl}$. Achieving this requires both an enhancement of the spectral tilt and enough suppression of the tensor-to-scalar ratio. Our results here are reminiscent of the fact that a slow-roll natural inflation model with $f\sim M_\mathrm{pl}$ can be easily realized in the presence of non-minimal derivative couplings \cite{Germani:2010hd}. In any case, we clearly see that the inflationary dynamics and the $n_s$-$r$ curves are sensitive to the $RT$ mixing term in the $f(R,T)$ gravity. The results shown here open a novel window for the scrutiny of inflation models in the $f(R,T)$ gravity with $RT$ mixing terms. 

We have to emphasize that the quantitative results presented in this paper rely strongly on the slow-roll approximations and the validity of the Einstein frame representations for the field equations. This assumption would not be valid near the end of inflation where the time derivatives of the inflaton are no longer negligible. We have numerically confirmed that, for all the cases considered in this paper, the absolute values of slow-roll parameters increase very rapidly (in terms of the inflaton value) near the end of inflation. Therefore, the slow-roll approximation is valid during the inflation and our calculations are expected to capture the qualitative behavior of the predictions in the parameter space. The calculations of the exact $n_s$-$r$ values in the presence of the $RT$ mixing term require further investigations and we will leave them to future work.  

\section*{Acknowledgement} 
CYC is supported by the Institute of Physics of Academia Sinica. YR is supported by the Natural Science Foundation of Xinjiang Uyghur Autonomous Region of China under grant No.~2022D01C52 and by the Doctoral Program of Tian Chi Foundation of  Xinjiang Uyghur Autonomous Region of China under grant No.~TCBS202128. XYZ is supported in part by the National Natural Science Foundation of China under grant No.~11835013 and by the Key Research Program of the Chinese Academy of Sciences under grant No. XDPB15.

\bibliographystyle{JHEP}
\bibliography{bibliography}

\end{document}